\documentclass[a4paper,12pt]{article}
\usepackage{jcappub} 

\newcommand{\Eq}[1]{eq.~\eqref{#1}}
\newcommand{\chirpM}{\mathcal{M}}
\newcommand{\dif}{\mathrm{d}}

\title{Estimating galactic foreground with the population of resolved galactic binaries}

\author[a,b]{Yang Jiang}
\author[a,b,c]{and Qing-Guo Huang}

\affiliation[a]{School of Fundamental Physics and Mathematical Sciences, \\
  Hangzhou Institute for Advanced Study, UCAS, Hangzhou 310024, China}
\affiliation[b]{School of Physical Sciences, \\
    University of Chinese Academy of Sciences, 
    No. 19A Yuquan Road, Beijing 100049, China}
\affiliation[c]{Institute of Theoretical Physics, \\
  Chinese Academy of Sciences,Beijing 100190, China}

\emailAdd{jiangy@ucas.ac.cn}
\emailAdd{huangqg@itp.ac.cn}

\abstract{The stochastic gravitational wave background (SGWB) in the millihertz band constitutes a primary science target for future space-based interferometers. The detection of such a signal, however, poses substantial data-processing challenges, not least due to the overwhelming contribution from numerous compact binaries in our Galaxy. Only a fraction of these binaries are sufficiently bright to be resolved as individual sources. The superposition of gravitational waves from the inspiral phases of the unresolved majority gives rise to a confusion foreground that must be accurately estimated. In this work, we model this foreground as an anisotropic SGWB and derive the variation of detector response intensity by analyzing the spatial distribution of binary systems. We further examine the distribution of resolved systems to assess whether they are representative of the total population. Finally, we perform parameter estimation to an injected stochastic background using the modeled foreground within the Taiji Data Challenge II. Our results demonstrate that a foreground description based on the population properties of resolved Galactic binaries can yield preliminarily viable estimates.
}

\begin{document}
\maketitle
\flushbottom

\section{Introduction}
Space-based gravitational-wave (GW) detectors, including the Laser Interferometer Space Antenna (LISA)~\cite{Baker:2019nia,LISA:2024hlh}, Taiji~\cite{10.1093/nsr/nwx116,10.1093/ptep/ptaa083}, and TianQin~\cite{Luo:2025sos}, will open the millihertz GW window in the coming decade. Operating in space avoids the seismic and gravity-gradient noise that limits ground-based interferometers, enabling the observation of massive black hole binaries~\cite{Li_2022}, extreme mass-ratio inspirals~\cite{Jonathan_2004}, Galactic binaries (GBs)~\cite{10.1093/mnrasl/slab003}, and potentially stochastic gravitational-wave backgrounds (SGWBs) of cosmological origin~\cite{Nishizawa:2011eq,Fumagalli_2022,Yu:2023lmo}. Among these sources, compact GBs are expected to dominate the millihertz band. Population synthesis predicts approximately $\mathcal{O}(10^7)$ binaries emitting quasi-monochromatic GWs in the Miky Way, of which only about $\mathcal{O}(10^4)$ can be individually resolved~\cite{Ruiter_2010}. The remaining unresolved binaries form a persistent Galactic confusion foreground that limits the sensitivity to detect other GW signals, making its accurate characterization an essential component of future data analysis.

A historical approach for estimating the Galactic foreground is to iteratively subtract individually resolved binaries above a chosen signal-to-noise (SNR) threshold and fit the residual with an empirical spectral model~\cite{PhysRevD.73.122001,Karnesis:2021tsh}. 
Although this approach provides a practical description of the residual spectrum, it relies on assumptions that are difficult to satisfy in realistic analyses. In particular, the subtraction procedure usually assumes perfectly known source parameters, whereas parameter estimation uncertainties inevitably leave residual signals in the data. It has been shown that these residuals modify the resulting confusion foreground level~\cite{Robson_2017}. Furthermore, empirical spectral models generally require prior knowledge of the instrumental noise and are often calibrated for specific datasets, limiting their applicability to realistic observations. Consequently, recent prototype global-fitting analyses~\cite{Littenberg:2023xpl} estimate the combined instrumental noise and confusion foreground without adopting a parameterized foreground model.

Moreover, describing the foreground solely by a power spectrum is incomplete. Because GBs are distributed anisotropically throughout the Milky Way and space-based detectors continuously change their orientation during orbital motion, the foreground exhibits a time-dependent amplitude modulation.
Existing spectral approaches largely adopts stationary assumption and neglect this behavior~\cite{Robson:2018ifk,Littenberg:2023xpl,Caprini:2024hue,Dimitriou:2025bvq}.
A promising alternative was recently proposed in \cite{Criswell:2024hfn}, where Galactic foreground is treated as an anisotropic SGWB whose detector response is determined by the spatial distribution of GBs. Since the large population of resolved Galactic binaries (RGBs) provides direct observational information on the Galactic structure~\cite{PhysRevD.89.022001}, this framework suggests that the unresolved foreground can be inferred from the RGB population itself, providing a physical connection between the observable and unresolved binary populations rather than relying on phenomenological fitting.

In this work, we present the first validation of this framework using realistic simulations of the Taiji Data Challenge II (TDC-II)~\cite{Du:2025xdq}. We first examine whether the resolved and unresolved binary populations share consistent distributions of intrinsic and spatial parameters, thereby testing the assumptions underlying the foreground model. Subsequently, we construct the anisotropic detector response based on the RGB population, estimate the amplitude of the unresolved Galactic foreground, and finally perform a search for an injected isotropic SGWB signal, illustrating the magnitude of the estimation error inherent to this scheme.

\section{Foreground and GB population}
\label{sec:fg}
The frequency evolution of GB is relatively slow and we can use a simple waveform~\cite{PhysRevD.106.103001}. 
In the source frame, it is given by
\begin{equation}
  h_+(t) = A(1+\cos^2\iota)\cos\left[\varphi(t)\right],\quad
  h_\times (t) = 2A \cos\iota \sin\left[\varphi(t)\right],
\end{equation}
where $\iota$ is the inclination angle. The phase takes a Taylor expansion:
\begin{equation}
  \varphi(t)=2\pi\left(f_0t + \frac12\dot{f}_0t^2 + \frac16\ddot{f}_0t^3\right) + \varphi_0,\quad
  \ddot{f}_0 = \frac{11}3\frac{\dot{f}_0^2}{f_0}.
\end{equation}
In the subsequent discussion on the population distribution, we prefer to parameterize with chirp mass $\chirpM$. 
For systems evolving solely under gravity, 
\begin{equation}
  A = \frac{2(G\chirpM)^{5/3}(\pi f_0)^{2/3}}{c^4d_L},\quad
  \dot{f}_0 = \frac{96\pi^{8/3}(G\chirpM)^{5/3}f_0^{11/3}}{5c^5}.
\end{equation}

Although the signal from unresolved GBs (UGBs) is so strong that it is termed a ``foreground", 
it is intrinsically random and statistical methods should be adopted.
We will model this foreground with the convention of anisotropic SGWB.
For Gaussian, stationary and unpolarized SGWB, the quadratic expectation satisfies
\begin{equation}
  \langle h_A(f,\boldsymbol{n})h_{A^\prime}^*(f^\prime,\boldsymbol{n}^\prime) \rangle=
  \frac{1}{4}S_c(f,\boldsymbol{n})\delta_{AA^\prime}
  \delta(f-f^\prime)\delta^2(\boldsymbol{n},\boldsymbol{n}^\prime).
  \label{eq:quadSGWB}
\end{equation}
It must be noted, however, that these assumptions are not strictly valid for the foreground. 
For example,~\cite{Rosati:2024lcs} shows the presence of non-Gaussianity in residuals of global fitting pipeline
hinders accurate estimation of the SGWB.
The inspiral frequency of GB increases gradually for a system dominated by GW radiation.
However, for the space mission with lifetime of several years, stability may be a feasible assumption.
Therefore, we retain the above assumptions to simplify the data processing.

The foreground is a superposition of GWs from individual GBs. 
Hence, the left side of \Eq{eq:quadSGWB} 
is the sum of signals from all GB systems
\begin{equation}
  \langle h_A(f,\boldsymbol{n})h_{A^\prime}^*(f^\prime,\boldsymbol{n}^\prime) \rangle = 
  \langle \sum_{i,j}h_{A,i}(f;\boldsymbol\theta_i^-)h_{A^\prime,j}^*(f^\prime;\boldsymbol\theta_j^-)\delta^2(\boldsymbol{n},\boldsymbol{n}_i)\delta^2(\boldsymbol{n}^\prime,\boldsymbol{n}_j) \rangle,
\end{equation}
where $i,j$ label GBs and $\boldsymbol\theta_{i,j}$ denote their parameters.
Note that here we have explicitly present the dependence on direction angle $\boldsymbol{n}_{i,j}$ and put a minus superscript on $\boldsymbol{\theta}$ to denote the remaining parameters. If we treat the GBs as independent samples drawn from a certain distribution and replace the ensemble average with the expectation,
the cross correlation terms of $i\neq j$ vanish. And $S_c(f,\boldsymbol{n})$ is determined by the population distribution of $\boldsymbol{\theta}$. 
We divide the parameters into three categories:
\begin{itemize}
  \item Intrinsic: chirp mass $\chirpM$, initial frequency $f_0$.
  \item Positional: ecliptic coordinates $\boldsymbol{n}=\{\lambda, \beta\}$, luminosity distance $d_L$.
  \item Others: inclination $\iota$, initial phase $\varphi_0$, polarization $\psi$.
\end{itemize}
We first state that the distributions of $\iota$, $\varphi_0$ and $\psi$ are decoupled from the rest. In other words, the joint distribution factorizes as
\begin{equation}
  p(\chirpM,f_0,\boldsymbol{n},d_L)p(\iota)p(\varphi_0)p(\psi).
\end{equation}
The independence arises primarily from the data simulation procedure: In galactic dynamical simulations, $\{\iota,\,\varphi_0,\,\psi\}$ are not modelled self‐consistently but are instead drawn afterwards from hypothetical uniform distributions in $\cos\iota$, $\varphi_0$ and $\psi$.
There is also a physical justification. This factorization ensures that the background is unpolarized and the signal is uncorrelated across different directions. We outline the derivation of the formalism of \Eq{eq:quadSGWB} with these assumptions in appendix~\ref{sec:fg_spectrum}.

Another commonly adopted assumption for anisotropic SGWB is that
the angular distribution of power is independent of frequency, namely,
\begin{equation}
  S_c(f,\boldsymbol{n})=H(f)P(\boldsymbol{n}). 
  \label{eq:decompSh}
\end{equation}
For a GB, the radiated frequency of GW depends only on $\chirpM$ and $f_0$.
Eq.~\eqref{eq:decompSh} implies a certain independence between $\{\chirpM,f_0\}$ and $\boldsymbol n$. 
In fact, it requires that the weighted marginalization over $d_L$
\begin{equation}
    \int \frac{p(\chirpM,f_0,\boldsymbol{n},d_L)}{d_L^2}\,\dif d_L
\end{equation}
factorizes into $\{\chirpM,f_0\}$ and $\boldsymbol n$.
This condition may seem somewhat complex. 
However, we generally regard the validity  of \Eq{eq:decompSh} as independent of the choice of coordinate origin.
Furthermore, the distributions of RGB and UGB populations are definitely different:
high-frequency systems are relatively few in number and experience weaker interference, making them easier to detect.
If the location distribution depended significantly on frequency, it would become impossible to reliably infer the properties of the UGBs from the resolved ones.
Therefore, we impose a stronger condition:
\begin{equation}
    p(\chirpM,f_0,\boldsymbol{n},d_L) = p(\chirpM,f_0)p(\boldsymbol{n}, d_L).
    \label{eq:decompFreqPos}
\end{equation}
We emphasize that this is an approximation to simplify data processing and does not represent the real situation.

Then we show in appendix~\ref{sec:fg_spectrum} that
\begin{equation}
  \begin{gathered}
  P(\boldsymbol{n}) = \int \dif d_L\,\frac{p(\boldsymbol{n},d_L)}{d_L^2}, \\
  H(f) = \frac{8N}{5}\int\dif\chirpM\, \mathcal{A}^2(\chirpM,f)p(\chirpM,f),
  \end{gathered}
  \label{eq:HP}
\end{equation}
where $N$ is the total number of GBs and $\mathcal{A}=Ad_L$.

\section{TDC II datasets}
Just as the approach in~\cite{Jiang:2026lik}, we still use the Taiji Data Challenge II (TDC II) training datasets~\cite{Du:2025xdq} 2\_8 to validate the SGWB search algorithm. 
Here, test-mass acceleration (ACC) and optical metrology system (OMS) displacement noise are simulated and the power spectral densities (PSDs) are
\begin{equation}
\begin{aligned}
  S_{ij,\text{acc}}(f) &=  A_{ij,\text{acc}}^2
      \left[1 + \left(\frac{0.4\,\text{mHz}}{f}\right)^2 \right]
      \left[1 + \left(\frac{f}{8\,\text{mHz}} \right)^4 \right]\left(\frac1{2\pi fc}\right)^2, \\
  S_{ij,\text{oms}}(f) &= A_{ij,\text{oms}}^2
        \left[1 + \left(\frac{2\,\text{mHz}}{f}\right)^4 \right]\left(\frac{2\pi f}c\right)^2. \\
\end{aligned}
\end{equation}
$ij$ labels the movable optical sub-assemblies (MOSAs) in the constellation. The nominal amplitudes are $A_\text{acc}=3\,\text{fm}/\text{s}^2/\sqrt{\text{Hz}}$ and $A_\text{oms}=8\,\text{pm}/\sqrt{\text{Hz}}$,
while there are $20\%$ fluctuations around the nominal value for actual amplitudes.
The injected astrophysical SGWB is a power law spectrum:
\begin{equation}
  \Omega_\mathrm{gw}(f)=A_\mathrm{ap}\left(\frac{f}{1\,\mathrm{mHz}}\right)^{\gamma_\mathrm{ap}},
  \label{eq:Omega_ap}
\end{equation}
where $A_\mathrm{ap}=10^{-12}$ and $\gamma_\mathrm{ap}=2/3$.
More details about the datasets and our conventions can be found in~\cite{Jiang:2026lik}. 
In that work, we neglected the effect of foreground and  pre-subtracted it prior to the main analysis.
Here, we will consider the original dataset.
The GB simulation generates about $4\times10^7$ binary systems according to the population model in~\cite{Korol:2021pun},
and then iteratively subtracts signals with SNRs greater than $7$. This process identifies $18984$ GBs as resolvable, with the remaining residual treated as the confusion foreground.
The parameters of both RGBs and UGBs are also provided in TDC II datasets.
Therefore, We first examine the discrepancies between our previous assumptions regarding the population distribution and the simulated data.

\begin{figure}[ht]
  \centering
  \includegraphics[width=.49\columnwidth]{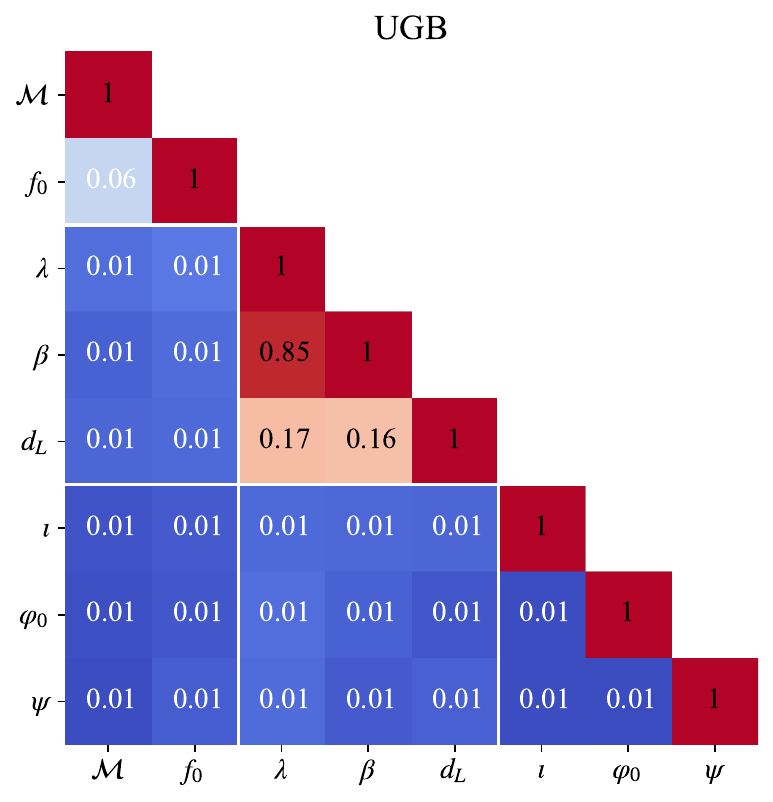}
  \includegraphics[width=.49\columnwidth]{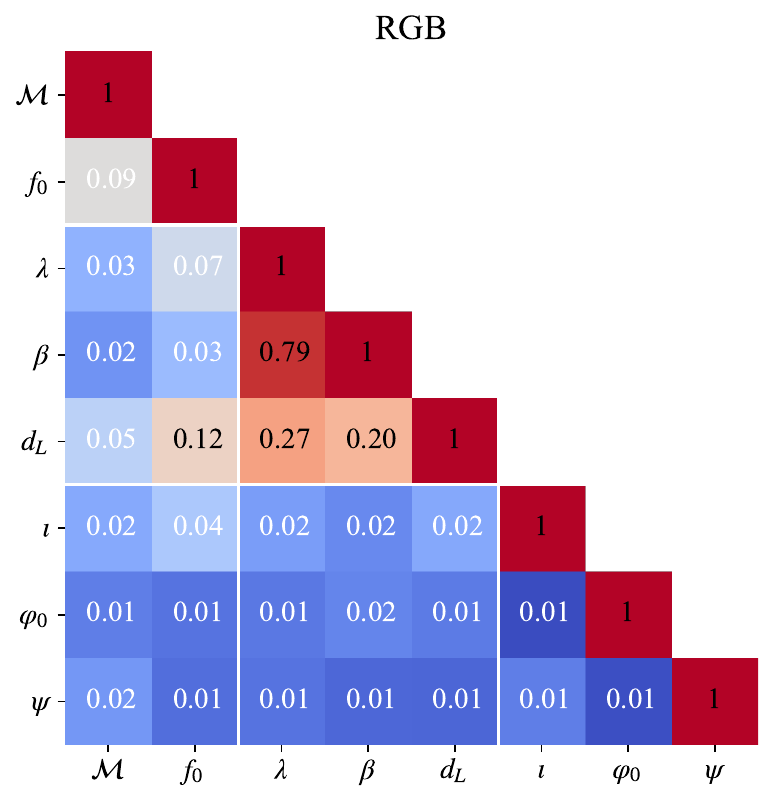}
  \caption{Distance correlations of parameters in GB systems. The left and right
  panels denote UGB and RGB datasets, respectively. For UGB dataset, we report the averaged correlation coefficient.}
  \label{fig:cor}
\end{figure}
We pay close attention to the independence between the parameters governing the position of GB and the radiated frequency,
as this impacts the validity of \Eq{eq:decompFreqPos} and \Eq{eq:decompSh}. 
We need an measure to quantify the correlation. The traditional Pearson’s coefficient is not suitable for this purpose.
Instead, we use the distance correlation~\cite{2008arXiv0803.4101S} as our metric.
Its value ranges from $0$ to $1$ and equals $0$ if and only if the two variables are independent. 
For RGBs, we compute the correlation with the injected values directly. 
While for UGBs, we randomly draw $100$ subsets containing the same number of binaries as in the resolved sets and then calculate the average correlation.
The results are shown in figure~\ref{fig:cor}. We have checked that the $2\sigma$ value of the correlation coefficient sampling does not exceed $0.001$. Therefore, the result reveals visible differences between the RGB and UGB populations. As expected,
the ecliptic azimuth and luminosity distance exhibits a strong correlation due to the anisotropic distribution of GBs. Meanwhile, the parameters $\{\iota,\varphi_0,\psi\}$ are known a prior to be  mutually independent. Their correlation coefficient are on the order of $\sim0.01$, which is comparable to the correlations between $\{\chirpM,f_0,\}$ and $\{\lambda,\beta,d_L\}$ in UGB dataset. We therefore conclude that the correlation between positional and frequency parameters is relatively weak for UGB dataset. But the dependence does appears to be present in RGB datasets.

In a realistic data analysis pipeline, the parameters of UGBs are unknown.
A possible feasible approach is to use the positional distribution of RGBs to approximate the unresolved ones.
Therefore, we also assess the differences between the two populations further.
\begin{figure}[ht]
  \centering
  \includegraphics[width=.9\columnwidth]{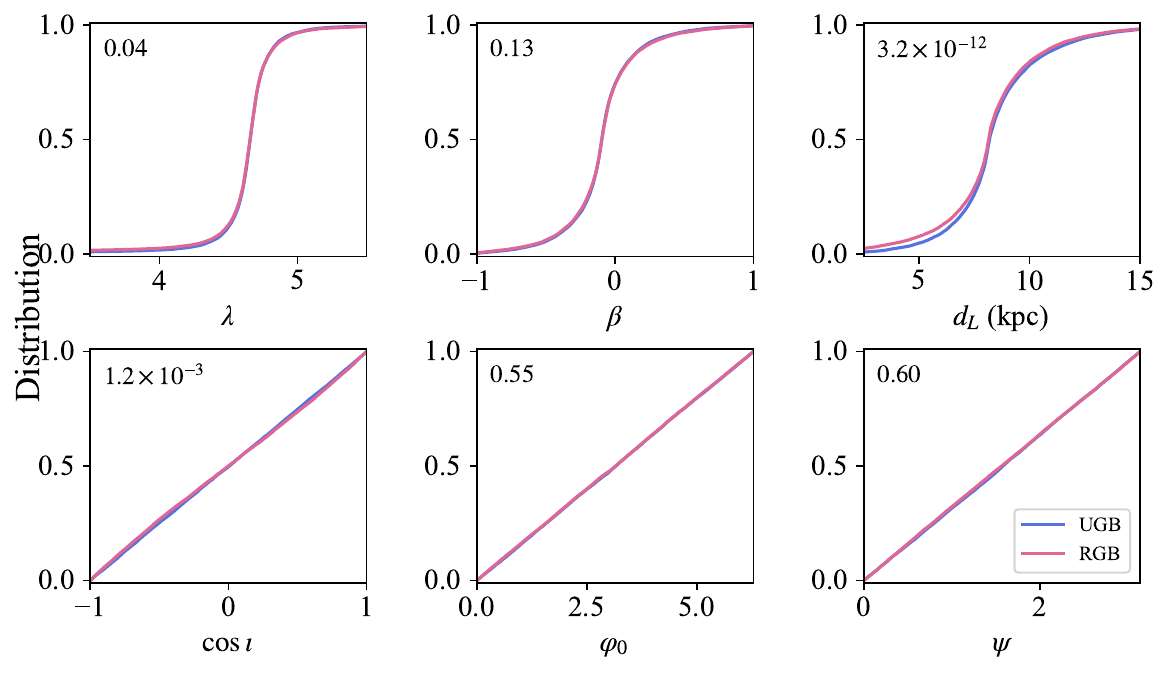}
  \caption{Distribution functions of parameters in GB system. The blue lines represent
  distributions of UGB dataset while the red lines represent RGB dataset. 
  $p$-values are marked in the upper left corner.}
  \label{fig:compdist}
\end{figure}
Figure~\ref{fig:compdist} shows the cumulative distributions of GB parameters.
At first glance, there appears to be little difference in the distribution of resolved and unresolved datasets. 
However, we further calculate the $p$-values using Kolmogorov-Smirnov test.
The $p$-value represents the possibility that two sampling sets are drawn from the same distribution.
Many parameters yield very low p-values, indicating that they are drawn from different distributions.
This is particularly evident for the luminosity distance $d_L$ and inclination angle $\iota$, for which theoretical explanations exist:
A large value of $\cos\iota$ leads to large amplitude of GB waveform.
Closer sources are more easily detectable and  thus tend to be classified as RGBs.
The existence of this selection effect complicates the issue and
hierarchical Bayesian inference~\cite{Toubiana:2026yml,Criswell2026} appears necessary.
However, we emphasize that although there are visible correlation and the distributions do not pass the identity test, 
this does not necessarily imply that they are also visible to the detector. 
We would like to verify whether there are observable bias with a simulation.

\section{Search for SGWB}
We begin by applying the time-delay interferometry (TDI) technique to the original signal.
We work with the $AET$ TDI combinations~\cite{PhysRevD.66.122002}. The cross-spectrum between two channels is
\begin{equation}
  \langle \tilde{s}_I(f)\tilde{s}_J^*(f^\prime)\rangle=\frac12\delta(f-f^\prime)\left[N_{IJ}(f)+\Gamma_{IJ}(f)S_h(f) +
  \int\dif^2\boldsymbol{n}\,\gamma(f,\boldsymbol{n})S_c(f,\boldsymbol{n}) \right],
  \label{eq:correlation}
\end{equation}
where $\tilde{s}_{I,J}(f)$ is the Fourier transform of signal. $N_{IJ}$ denotes the correlation of instrumental noise. $S_h(f)$ is the PSD of injected isotropic SGWB.
The isotropic overlap reduction function is
\begin{equation}
  \Gamma_{IJ}(f) = \frac1{4\pi}\int\dif^2\boldsymbol{n}\,\gamma(f,\boldsymbol{n}).
\end{equation}
To evaluate the angular integral, we employ a pixelization scheme on the sphere using the \texttt{healpy} package~\cite{Zonca2019,2005ApJ622759G},  which discretizes the integral into a sum over sky pixels. 
\texttt{nside} of $18$ is adopted in our work.

The angular distribution $P(\boldsymbol{n})$ is fixed to previous result of RGB population analysis.
This is achieved by counting the number of sources in each pixel and averaging them with $d_L^{-2}$ as the weight.
However, a problem arises here. 
Since we do not introduce a prior model for the structure of Milky Way, 
the sky map is obtained directly with the position of simulated GBs. 
In some sky regions where the number of GBs is insufficient, 
there is a significant deviation between sampling and statistical average. 
Specifically, these directions may exhibit excessively bright noise.
Therefore, we have set a threshold of $5$ GBs:
Sky regions containing fewer than this number of GBs are not be involved in the calculation of foreground.

\begin{figure}[ht]
  \centering
  \includegraphics[width=.7\columnwidth]{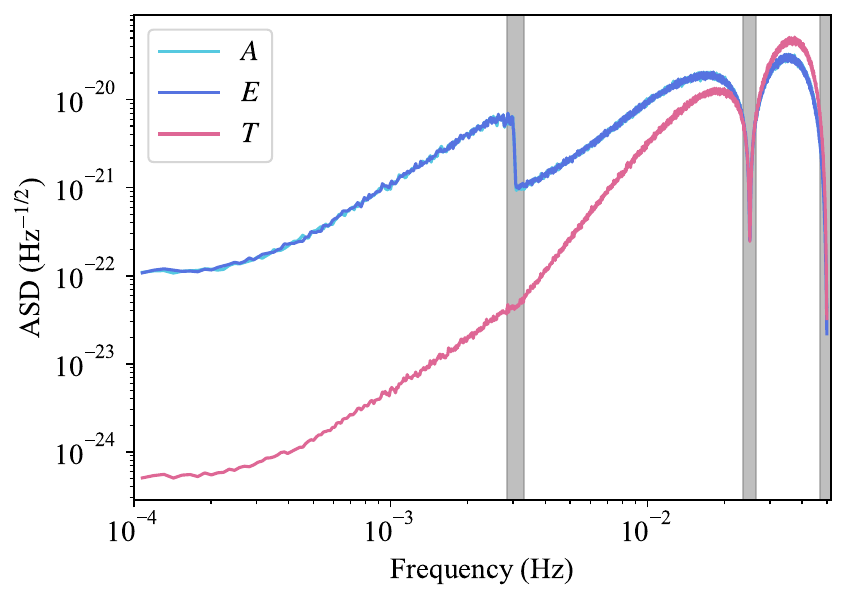}
  \caption{Averaged ASD of TDC II simulated data in $A,E,T$ channels. 
  Grey shaded regions denote frequencies notched out in the analysis.}
  \label{fig:asd}
\end{figure}

$H(f)$ and the parameters of the isotropic SGWB spectrum will be inferred from the simulated data within a Bayesian framework.
Owing to variations of armlength and orientation of detectors, the observed TDI data are non-stationary.
Therefore, we split the data $s$ into segments, each lasting $5$ days, and transform each segment into frequency domain via
\begin{equation}
    \tilde{s}_{l,I}[k] = \delta t\sum_{j=0}^{N -1}s_{l,I}[j]\mathrm{e}^{-\mathrm{i}2\pi jk/N_l},
    \label{eq:fft}
\end{equation}
where $l$ indexes the segments. $\delta t$ is the cadence and $N_l$ is the number of samples per segments. We use square brackets to denote discrete samplings with the correspondence
\begin{equation}
\begin{gathered}
    g[j]\sim g(t_j),\quad t_j = j\delta t;\\
    \tilde{g}[k]\sim \tilde{g}(f_k),\quad f_k=k\delta f.
\end{gathered}
\end{equation}
for any continuous signal. We assume that the statistical properties are stationary within each segment. The likelihood is
\begin{equation}
  \begin{gathered}
    \ln p(s|C) = -\sum_{l,k}\left[\ln \left|C_l(f_k)\right| + \frac{2}{T_l}\tilde{s}_l^\dagger(f_k)C_l^{-1}(f_k)\tilde{s}_l(f_k)\right],\\
    \tilde{s}_l(f_k) = \begin{bmatrix}\tilde{s}_{l,A}(f_k) & \tilde{s}_{l,E}(f_k) & \tilde{s}_{l,T}(f_k)\end{bmatrix}^T,
  \end{gathered}
  \label{eq:likelihood}
\end{equation}
up to an additive constant. $T_l$ is the duration of a segment. The components of covariance $C_l$ are given by \Eq{eq:correlation}.
We show the average amplitude spectral density (ASD) of $\tilde{s}_l(f)$ in figure~\ref{fig:asd}.

The analyzed frequency band ranges from $0.1$ mHz to $0.05$ Hz, 
with some frequency bins removed due to the degenerate sensitivity of TDI.
Since we have no prior information about how to parameterize $H(f)$, 
this spectrum is fitted using Akima interpolation~\cite{10.1145/321607.321609}. 
The number and positions of the knots are determined via trans-dimensional Markov chain Monte Carlo (MCMC).
This is performed by reverse jump MCMC sampler in \texttt{Eryn}~\cite{Karnesis2023,michael_katz_2023_7705496}.
The estimated band for $H(f)$ also starts from $0.1$ mHz.
In the frequency band exceeding about $3$ mHz, the sharp reduction in the number of UGBs
leads to a rapid decrease of the foreground spectrum.
When the number of sources is insufficient, Gaussianity may no longer hold.
Consequently, we set the the upper frequency band to be $2.85$ mHz and several portion of the subsequent frequency bins are cut off.
For $12$ amplitudes of instrumental noise, we set uniform priors spanning $\pm25\%$ fluctuations around the nominal values.
For amplitude and power index in \Eq{eq:Omega_ap}, we set uniform priors in $[10^{-14},\, 5\times10^{-11}]$ and $[-1, 2]$, respectively.

\section{Results}
\begin{figure}[ht]
  \centering
  \includegraphics[width=.7\columnwidth]{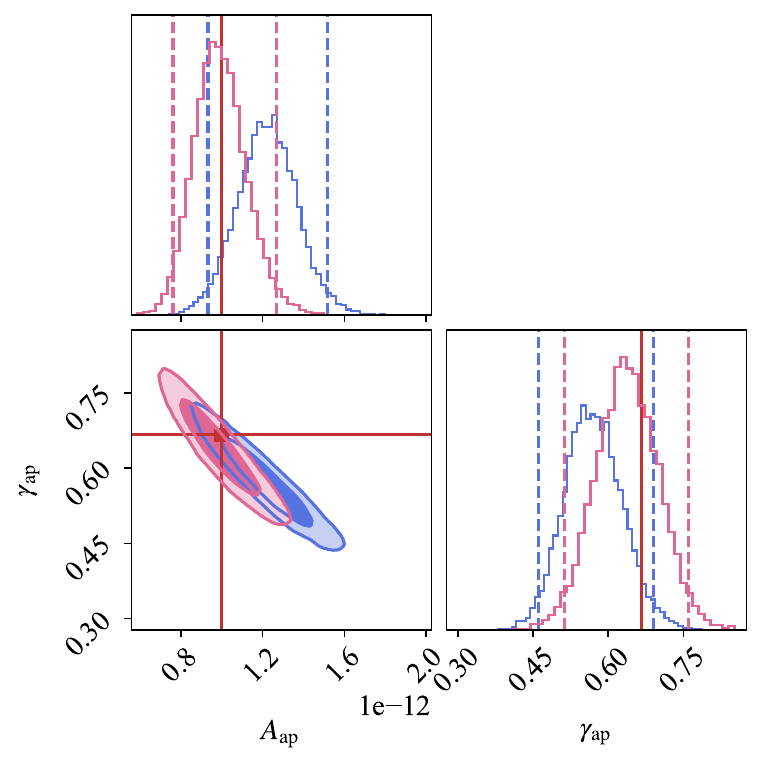}
  \caption{Posterior distributions for the parameters of the astrophysical SGWB spectrum.
  The red lines denote injected values while the vertical dashed lines indicate $95\%$ CLs.
  The shaded regions represent $68\%$ and $95\%$ contours of CLs. The blue regions correspond to results obtained with UGB population,
  and the pink regions to results with RGB population.}
  \label{fig:corner}
\end{figure}

\begin{figure}[ht]
  \centering
  \includegraphics[width=.7\columnwidth]{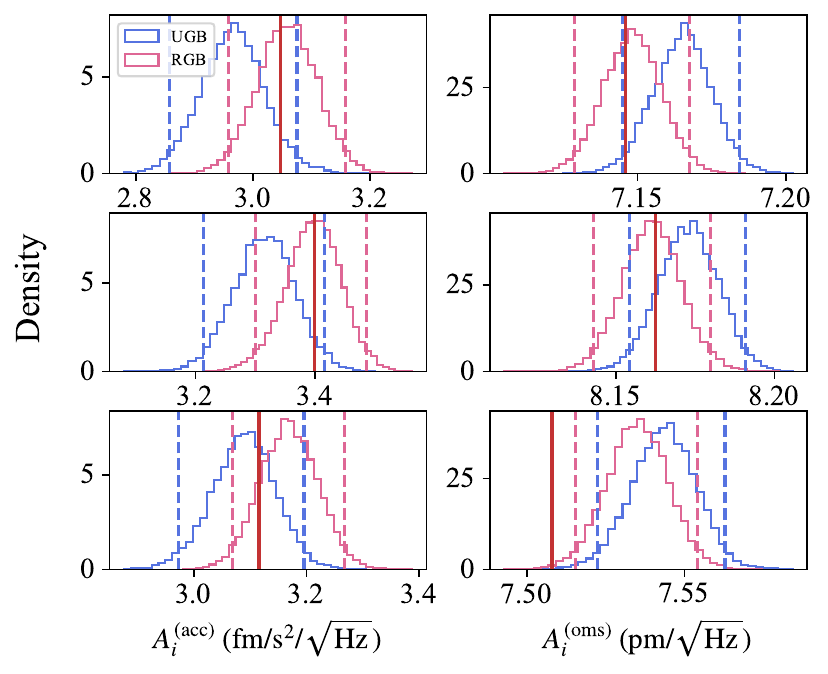}
  \caption{Posterior distributions for ACC and OMS noise amplitudes,
  arranged in ascending order of spacecraft index $i$ from top to bottom.
  The red lines denote injected values while the vertical dashed lines show the $2\sigma$ CLs.}
  \label{fig:noise}
\end{figure}

Before estimating the distribution using RGBs, we perform the same procedure with UGBs
by drawing $4\times10^7$ source in unresolved datasets.
Estimating the population distribution with the given positions and then carrying out Bayesian analysis.
The posterior distribution of parameters in the astrophysical SGWB spectrum \Eq{eq:Omega_ap} is shown in blue in figure~\ref{fig:corner}.
The injected values is located at the edge of $68\%$ confidence level (CL) of the posterior distribution.
For the amplitude of noise, we define $A_i=(A_{ij}+A_{ik})/2$ and express the result using this quantity to
mitigate degeneracy between the parameters~\cite{PhysRevD.82.022002}.
The posterior, shown in figure \ref{fig:noise}, 
depicts the distribution of the noise amplitude.
Most injected values can be recovered with $2\sigma$ accuracy.
Next, the result with RGB population analysis are also demonstrated in the figures but with pink regions or lines. 
The deviations to the injected values are similar to the results
from UGBs.
It is not evident that there are biases in the results of parameter recovery
except for a amplitude of OMS noise in the bottom line of figure \ref{fig:noise}.

The restoration of the spectrum $H(f)$ for the foreground is also a meaningful result. 
However, treating the foreground in the same way as the background is a method we choose artificially.
There is no corresponding injection value in the dataset for a comparison. 
We decide to include this result in the appendix \ref{sec:recov_fg}.

\section{Conclusions}
In this work, we have searched for the SGWB amidst the confusion foreground generated by the superposition of 
GWs from GBs. 
Our method decomposes the foreground into two independent parts: 
the spectrum in the frequency domain and the modulation effect in the time domain.
For the former, we employ a flexible interpolation and trans-dimensional sampling to fit it. 
For the latter, we first synthesize the positional information of binary star systems in the dataset 
to derive their distribution characteristics,
and then use the response pattern of detectors to calculate this modulation effect.

Most unknown parameters can be recovered with a CL of $2\sigma$.
This error is not significant but can be perceived to some extent.
This is likely related to the assumptions we made when describing the foreground spectrum.
Some information is smoothed out during the process of transforming individual system parameters into statistical distributions. 
When calculating the energy spectrum, we assume Gaussianity and independence among different sources. 
In regions with a small number of sources, this assumption may not hold. 
From an observational perspective, 
the operational average of background PSD can only be achieved by averaging data taken at different times. 
The expectation of a probability distribution, on the other hand, assumes a ensemble average across multiple galaxy systems, 
and the two are not equivalent. 
Therefore, deviations in the energy spectrum calculated from statistical properties are understandable. 
Furthermore, due to the selection effects of detection,  
the distribution characteristics of the resolvable dataset may not even represent the distribution model of unknown sources accurately.

Nevertheless, the current approach may still retain practical value. 
In a complete data analysis process, the results of parameter estimation of resolved GB systems also contain errors.
Hence, the spatial distribution model should also be accompanied by uncertainty, 
and we need to compare this uncertainty with the detection capability of the detector to draw more accurate conclusions. 
Then, we will consider whether it is really necessary to take the selection effect into account.
This requires us to further establish a process that combines GB identification with foreground estimation.

\appendix

\section{Spectrum of foreground}
\label{sec:fg_spectrum}
The method we use to relate the energy spectrum and population distribution is taken from \cite{Belgacem2025}, 
which focuses on coalescing compact binaries.
We make several modifications to adapt it to continuous GB signals.
In section \ref{sec:fg}, we have explained that
\begin{multline}
  \langle h_A(f,\boldsymbol{n})h_{A^\prime}^*(f^\prime,\boldsymbol{n}^\prime) \rangle \sim
  N \int\dif\chirpM\dif f_0\, p(\chirpM, f_0) 
  \int\dif^2\boldsymbol{n}_i\dif d_L\, p(\boldsymbol{n}_i,d_L)
  \delta^2(\boldsymbol{n},\boldsymbol{n}_i)\delta^2(\boldsymbol{n}^\prime,\boldsymbol{n}_i) \\
  \int\dif\iota\,p(\iota)
  \int\dif\varphi_0\,p(\varphi_0) 
  \int\dif\psi\,p(\psi) h_A(f)h_{A^\prime}^*(f^\prime).
\end{multline}
Note that $h_A$ on the right side represents the waveform of a single system and
we have omitted the integrated variables $\boldsymbol{\theta}_i^-$.
We have also omitted most of index $i$ from parameters for simplicity.

We can explicitly express the dependence of waveform on the polarization angle $\psi$ as
\begin{equation}
  \sum_A R_{AB}(2\psi)h_B(f).
\end{equation}
That is, we redefine $h_B$ to denote the particular waveform of $\psi=0$.
We will frequently use similar notations in subsequent derivations to ease complex labels.
The matrix $R_{AB}$ is
\begin{equation}
  R_{AB}(2\psi) = 
  \begin{bmatrix}
    \cos2\psi & -\sin2\psi \\ \sin2\psi & \cos2\psi
  \end{bmatrix}_{AB} = 
  \delta_{AB}\cos2\psi - \mathrm{i}\sigma_{AB}^2\sin2\psi,
\end{equation}
where $\sigma^2_{AB}$ is the second Pauli matrix.
Marginalize over $\psi$ gives
\begin{equation}
  \frac12\delta_{AA^\prime}\sum_B h_B(f)h_B^*(f^\prime) + \frac{\mathrm{i}}2\sigma_{AA^\prime}^2
  \left[h_+(f)h_\times^*(f^\prime) - h_\times(f) h_+^*(f^\prime)\right]
  \label{eq:margpsi}
\end{equation}

Since we assume that the foreground is stationary, we neglect the derivations of initial frequency
and treat the GW signal as monochromatic. The waveform in frequency domain is
\begin{equation}
  h_+(f) = \frac{\mathcal{A}}{d_L}\delta(f-f_0)\frac{1+\cos^2\iota}{2}\mathrm{e}^{\mathrm{i}\varphi_0},\quad
  h_\times(f) = \frac{\mathcal{A}}{d_L}\delta(f-f_0)\cos\iota\mathrm{e}^{\mathrm{i}(\varphi_0-\frac{\pi}2)},
\end{equation}
where we have ignored the negative frequencies.
Because waveform appeared in \Eq{eq:margpsi} always couples with its complex conjugation, 
$\varphi_0$ cancels out, and marginalizing over it leaves the result unchanged.

When marginalize over $\iota$, the term containing Pauli matrix in \Eq{eq:margpsi} vanishes due to the integral of an odd function and 
the first item yield a coefficient
\begin{equation}
  \frac12\int_{-1}^1\dif(\cos\iota)\, \left[\left(\frac{1+\cos^2\iota}2\right)^2 + \cos^2\iota\right]
  = \frac45.
\end{equation}
Thus the integrand becomes
\begin{equation}
  \frac{2}{5}\delta_{AA^\prime}\delta(f-f_0)\delta(f^\prime-f_0)\left(\frac{\mathcal{A}}{d_L}\right)^2.
\end{equation}

Using
\begin{equation}
  \delta^2(\boldsymbol{n},\boldsymbol{n}_i)\delta^2(\boldsymbol{n}^\prime,\boldsymbol{n}_i) =
  \delta^2(\boldsymbol{n},\boldsymbol{n}^\prime)\delta^2(\boldsymbol{n},\boldsymbol{n}_i),
\end{equation}
and
\begin{equation}
  \delta(f-f_0)\delta(f^\prime-f_0) = \delta(f-f^\prime)\delta(f-f_0).
\end{equation}
We complete the integration and the final result is
\begin{equation}
  \frac25\delta_{AA^\prime}\delta(f-f^\prime)\delta^2(\boldsymbol{n},\boldsymbol{n}^\prime)
  \int\dif\chirpM\, \mathcal{A}^2(\chirpM,f)p(\chirpM,f)
  \int \dif d_L\,\frac{p(\boldsymbol{n},d_L)}{d_L^2}.
\end{equation}
Comparing with \Eq{eq:quadSGWB} leads to \Eq{eq:HP}.

\section{Recovered spectrum of foreground}
\label{sec:recov_fg}
\begin{figure}[ht]
  \centering
  \includegraphics[width=.7\columnwidth]{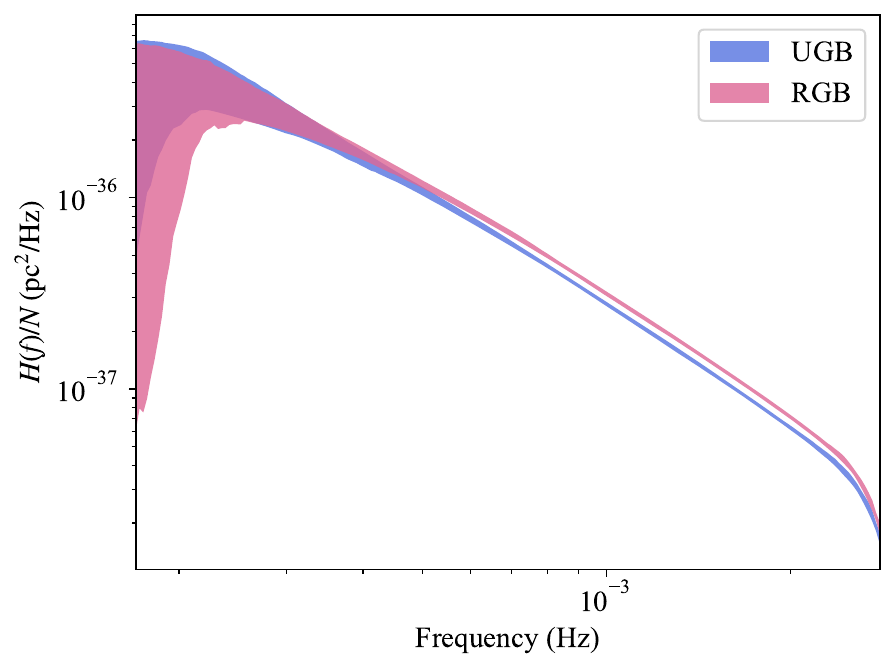}
  \caption{Recovered energy spectrum $H(f)$ of confusion foreground with RGB and UGB population estimation. 
  The shaded region represent $95\%$ CL.
}
  \label{fig:Hf}
\end{figure}

Here, we present the recovered frequency spectrum $H(f)$ of the confusion foreground.
This is achieved by translating the posterior knots into the spectrum and calculate CLs at selected frequencies.
The result is shown in figure \ref{fig:Hf}.
We notice an inconsistency between the results within the sensitive band of the detector. 
This discrepancy arises because we manually excluded some GBs when calculating the spatial distributions. 
This operation is only performed on the RGB dataset, 
resulting in a slightly underestimated value of $P(\boldsymbol{n})$.
As a compensation, the value of $H(f)$ appears to be larger. 
This is a necessary sacrifice trade-off to alleviate sampling errors.

\acknowledgments

YJ is supported by the China Postdoctoral Science Foundation under Grant Number 2025M783376. 
QGH is supported by the grants from NSFC (Grant No.~12547110, 12475065, 12447101) 
and the China Manned Space Program with grant no. CMS-CSST-2025-A01.

\bibliographystyle{JHEP}
\bibliography{refs.bib}

\end{document}